\documentclass[12pt]{article}
\usepackage{amsmath}
\usepackage{amssymb}
\usepackage{cite}
\usepackage{float}
\usepackage{latexsym}
\usepackage{graphicx}
\newcommand{\fft}[2]{{\frac{#1}{#2}}}
\newcommand{\ft}[2]{{\textstyle\frac{#1}{#2}}}

\newcommand{\be}{\begin{equation}}
\newcommand{\ee}{\end{equation}}
\newcommand{\bea}{\begin{eqnarray}}
\newcommand{\eea}{\end{eqnarray}}
\newcommand{\dx}{\partial X}
\newcommand{\dbx}{\bar{\partial} X}
\newcommand{\nn}{\nonumber}
\begin{document}
%%%%%%%%%%%%%%%%%%%%%%%%%%%%%%
\begin{titlepage}

\begin{flushright}
MCTP--05--102
\end{flushright}

\vspace*{1cm}

\begin{center}
{\bf\Large Spinning strings in AdS$_5\times S^5$:\\[8pt]
A worldsheet perspective}

\vspace*{1cm}

{Benjamin A. Burrington and James T. Liu}

\vspace*{.5cm}

{\it Michigan Center for Theoretical Physics, Randall Laboratory of Physics,\\
The University of Michigan, Ann Arbor, MI 48109--1040}

\vspace*{.5cm}

{E-mail: \tt bburring@umich.edu, jimliu@umich.edu}
\end{center}

\vspace*{1cm}

\begin{abstract}
We examine the leading Regge string states relevant for
semi-classical spinning string solutions.  Using elementary RNS techniques,
quadratic terms in an effective Lagrangian are constructed which describe
massive NSNS strings in a space-time with five-form flux.
We then examine the specific case of AdS$_5\times S^5$, finding the dependence of
AdS ``energy'' ($E_0$) on spin in AdS ($S$), spin on the sphere ($J$), and
orbital angular momentum on the sphere ($\nabla_a \nabla^a$).
\end{abstract}
\end{titlepage}

%%%%%%%%%%%%%%%%%%%%%%%%%%%%%%

\section{Introduction}

Many of the most remarkable developments in M-theory may be traced to
the study of various dualities.  Although there is as yet no complete
understanding of M-theory in its entirety, it is nevertheless clear that many
aspects of the theory are very well captured through dualities of various sorts.
In particular, the notion of AdS/CFT duality only achieves its full
potential when connecting the various pieces---the open string worldsheet,
D-brane worldvolume actions, supergravity $p$-brane solutions, low energy
effective theories, {\it etc.}---into a complete whole.  In this manner,
AdS/CFT duality becomes more than a general statement on isometries, but
can be applied to very specific cases such as the duality between
$\mathcal N=4$ super-Yang Mills theory and strings on AdS$_5\times S^5$.

Even within this single context of $\mathcal N=4$ super-Yang Mills, different
methods have been applied to enlarge the regime of investigation.  The
gauge/gravity duality itself relates the 't~Hooft coupling of the gauge
theory, $\lambda\equiv g_{\rm YM}^2N$, to the string theory through
$\sqrt\lambda=L^2/\alpha'$ where $L$ is the `radius' of AdS$_5$ (or
equivalently $S^5$).  In general, perturbative results on the gauge theory
side may be trusted in the limit $\lambda\ll1$, while the supergravity
limit yields the opposite regime, $\lambda\gg1$.  Thus explicit tests of
AdS/CFT in this context have been hindered by the strong/weak coupling
nature of the duality.  Fortunately, techniques have been devise to move
beyond the supergravity limit, where long strings start playing an important
r\^ole.  In particular, Berenstein, Maldacena and Nastase
\cite{Berenstein:2002jq} were able to
explore the large $R$-charge sector by taking a Penrose limit of AdS$_5\times
S^5$.  The resulting expansion is then controlled by $\tilde\lambda=\lambda/J^2$
where $J^2=J_1^2+J_2^2+J_3^2$ is the sum of the three commuting $R$-charges
resulting from angular momentum on $S^5$.  The resulting string admits an
exact treatment in the Green-Schwarz formalism in the plane-wave background
\cite{Metsaev:2002re,Russo:2002rq}.

At the same time, one may also explore the far from BPS sector of strings
with large spin in AdS$_5$.  This was demonstrated by Gubser, Klebanov and
Polyakov \cite{gkp}, who took a semi-classical spinning string in AdS$_5$ on
the leading Regge trajectory.  For short strings (spin $S\ll\sqrt\lambda$),
they obtained $E\approx ML$ where the closed string mass is given by
$\alpha'M^2=2(S-2)$.  On the other hand, this relation turns over to
$E\approx S+(\sqrt\lambda/\pi)\ln(S/\sqrt\lambda)$ in the long string limit.
Many subsequent investigations have led to refinements of these results as
well as additional progress in obtaining anomalous dimension relations for
operators carrying various spin quantum numbers in AdS$_5$ and $R$-charge on $S^5$
\cite{Frolov:2002av,Tseytlin:2003ii,Tseytlin:2004xa,Park:2005ji,krt,kt,bdr,Frolov:2003xy,Frolov:2003qc,
Arutyunov:2003uj}.
Much of
the stringy analysis, however, have been classical or at most semi-classical
in nature, as quantization of the string worldsheet is rather unpleasant in
a RR background
\cite{Kallosh:1998nx,Metsaev:1998it,Kallosh:1998ji,Drukker:2000ep}
(except in special cases such as the Green-Schwarz string
in an plane-wave).

In this paper, we investigate the use of covariant worldsheet (RNS) methods
to study highly excited string states on the leading Regge trajectory in
the AdS$_5\times S^5$ background.  Our approach is to apply elementary
techniques to first covariantly quantize the IIB string in a ten-dimensional
flat background, and then to derive a Minkowski-space effective action
governing massive string states up to a particular mass level $n$ given
by $\alpha' M^2=4(n-1)$.  In this manner, we are in principle able to deduce
the interactions of any particular massive state with the massless
IIB fields ({\it i.e.}\ the metric and RR 5-form) as well as the
other states in the massive tower.  Although this effective action is
constructed perturbatively about a flat background, we note that general
covariance fixes its form, at least up to terms proportional to the equations
of motion (which vanish on-shell in the Minkowski background).
As a result, this effective action also governs the interactions of
massive string states with any background, and in particular the AdS$_5
\times S^5$ background with non-trivial metric and 5-form turned on.
Furthermore, any undetermined terms proportional to the equation of motion
computed in the Minkowski background will likewise vanish on the
AdS$_5\times S^5$ background (as both backgrounds satisfy identical
equations of motion).  In this manner, we are able to perturbatively
investigate the behavior of massive string states (which, for simplicity,
we take on the leading Regge trajectory) in AdS$_5\times S^5$ using
straightforward RNS worldsheet techniques.

In general, for an effective action to properly describe string states
up to a particular mass level $n$, it must contain all fields up to and
including the level of interest.  From an effective field theory point of
view, this corresponds to integrating out massive modes above a cutoff scale
set by $n$.  While this is in principle a straightforward task, the
enormous number of string fields would turn this into a hopelessly long
endeavor.  Thus, in practice, we focus only on the propagation of a single
massive field $\Phi$ on the leading Regge level interacting with the
background.  This leads to an effective (tree-level string) Lagrangian of
the schematic form
\begin{eqnarray}
e^{-1}\mathcal L&=&R-\ft1{2\cdot5!}F_{(5)}^2+\cdots\nonumber\\
&&-\ft12\nabla_\mu\Phi\nabla^\mu\Phi-\ft12M^2\Phi^2+\sum_{k\ge1}
(\alpha')^{k-1}\Phi\{R,F_{(5)}^2,\nabla^2\}^k\Phi+\cdots,
\label{eq:schemlag}
\end{eqnarray}
where we have only highlighted the terms of interest in the $\alpha'$
expansion.  In general, of course, the string expansion yields an effective
action expanded in both $\alpha'$ and $g_s$.  However, we content ourselves
with working in the large $N$ limit, corresponding to tree level on the
string side of the duality.

By omitting cubic and higher interactions of $\Phi$ with lower mass fields
in (\ref{eq:schemlag}), we are essentially postulating that $\Phi$ is
sufficiently
stable so that it may be effectively treated as a single particle state as
opposed to a broad resonance.  In the underlying string picture, this requires
that the closed string on the leading Regge trajectory has a sufficiently
narrow width.  At first, this appears a somewhat dangerous assumption, because
of the large number of stringy decay products.  However, many potential
decays turn out to be exponentially suppressed, and in particular it was
shown in Refs.~\cite{Iengo:2002tf,Iengo:2003ct} that (at least in a
Minkowski background) such leading Regge states of mass $M$ have a lifetime
on the order $\tau=\mathcal O(\alpha'M)$.  Thus highly massive states are
in fact long lived, and it is therefore valid to only examine the particular
massive state $\Phi$ without bringing in the entire tower of string states
below mass $M$.  Semiclassical decay of strings in an AdS$_5 \times S^5$ background
were studied in \cite{Peeters:2004pt}, and the dual gauge theory phenomena
in \cite{Peeters:2005pb}.

In the perturbative expansion of the effective Lagrangian (\ref{eq:schemlag}),
the higher $\alpha'$ terms correspond to the interaction of a string of
finite extent with the background curvature and $5$-form flux.  From this
point of view, we may estimate the size of the $k$-th order term as
$(n\alpha'/L^2)^k$ where $n$ is the string level and $L$ is the radius of
AdS$_5$ (or $S^5$).  This estimate arises because the length of a highly
spinning string essentially grows as $\sqrt{n\alpha'}$.  As a result, validity
of the perturbative expansion requires that $L^2\gg n\alpha'$, or
equivalently that $S\ll\sqrt\lambda$, corresponding to short strings spinning
in AdS$_5$.

Technically we calculate only the first non-trivial interactions of the
massive state $\Phi$ with the background.  This requires a stringy 3-point
computation in the NSNS sector to obtain the $\Phi R\Phi$ interaction as
well as a 4-point computation with two Ramond-emission vertex operators to
obtain the $\Phi F_{(5)}^2\Phi$ term.  Although the computations are not
difficult, they are somewhat involved, especially when extracting the
$\Phi F_{(5)}^2\Phi$ contact term from the 4-point function.  Thus we have
also performed a check on our results by taking the massless graviton limit
$\Phi\to h_{\mu\nu}$ to verify that we reproduce the expected interactions
in the massless sector.

At this order, we are able to extract the first correction to the flat-space
Regge behavior.  For a state on the leading Regge trajectory spinning entirely
in AdS, we obtain $\alpha'\mu^2=2(S-2)-\alpha'S^2/2L^2$, which is valid so long
as the correction term is much less than the flat-space term $2(S-2)$.
(Here $\mu^2$ is related to mass in AdS, and will be properly defined below.)
Measuring the energy in AdS units results in the short
string Regge relation
\begin{equation}
E=2+\sqrt{\fft{2(S-2)L^2}{\alpha'}-\fft{S(S-2)}2+4+\cdots}
\approx\sqrt{2\sqrt\lambda S}\left(1-\fft{S}{8\sqrt\lambda}+\cdots\right),
\end{equation}
where the second expression is valid in the semi-classical regime
$1\ll S\ll\sqrt\lambda$.  The leading term is simply the flat-space Regge
behavior of short spinning strings, while the correction at least suggests
that the Regge trajectories flatten out in the limit of large spin.  Of
course, the actual $S\to\infty$ limit corresponds to the long string limit,
which is unattainable by this perturbative method.

In the next section, we give an overview of the method of extracting the
effective action (\ref{eq:schemlag}) from the string $S$-matrix and in
particular lay out the actual structure of the interaction Lagrangian.
Following that, in section 3 we compute the $\Phi$-$h_{\mu\nu}$-$\Phi$
three-point function and in section 4 we work out the
$\Phi$-$F_{(5)}$-$F_{(5)}$-$\Phi$ four-point function.  Technically, this
is the most involved step of the calculation, as it requires pole subtractions
to remove terms that may be factorized on three-point functions.
We then work out the implications for short-string Regge behavior and
conclude with a discussion of these results in section 5.

%%%%%%%%%%%%%%%%%%%%%%%%%%%%%%
\section{Extracting the effective action}

In string theory, the basic objects of the underlying CFT are vertex operators,
which encode the asymptotic states of the theory.  With these, one may compute
$S$-matrix elements corresponding to the scattering of on-shell string states.
In computing $n$-point S-matrix elements, however, one is faced with divergences
for certain values of external momenta.  These are interpreted as the exchange
of on shell degrees of freedom, with the divergences coming about from physical
poles in the propagators $\sim1/(p^2+m^2)$.  However, String theory (in a
first quantized approach) has no string creation or annihilation operators,
and so defining a ``string propagator'' is not possible.  Nevertheless,
by considering all $n$-point string scattering amplitudes, one may construct an
effective particle field theory description, with each vertex operator being
associated with a field in the theory.  The effective field theory may be
encoded in a Lagrangian whose $S$-matrix elements correctly reproduce the
string $S$-matrix.  In this description, the pole terms can be calculated,
and directly correspond to the divergences in the string $S$-matrix elements.
In practice, of course, one only calculates the first few scattering amplitudes,
and then appeals to further symmetry arguments to determine the effective
Lagrangian to the desired order.

As indicated above, we explore massive states on the leading Regge trajectory
in the NSNS sector of the IIB string.  For a given mass level $n$, these
closed string states have mass $\alpha'M^2=4(n-1)$ and carry spin $S=2n$
(in the flat ten-dimensional background).  For a closed bosonic string they
would be generated by vertex operators of the form
\begin{equation}
V(\zeta,k)=\zeta_{\mu_1\cdots\mu_{2n}}\partial X^{\mu_1}\cdots\partial X^{\mu_n}
\partial\overline X^{\mu_{n+1}}\cdots\partial\overline X^{\mu_{2n}}
e^{ik\cdot X},
\end{equation}
where the polarization tensor $\zeta_{\mu_1\cdots\mu_{2n}}$ is transverse and
trace-free.  The actual superconformal vertex operators are only slightly
more complicated, and will be presented below.  In the following two sections,
we compute the appropriate scattering amplitudes to find the effective field
theory couplings to the background curvature and 5-form (which are the only
massless IIB fields needed for the AdS$_5\times S^5$ background).

To describe the effective Lagrangian, we may associate a field
$\Phi^{\mu_1 \cdots \mu_{2n}}$ to the above leading Regge trajectory state.
In principle, much care is needed in writing out a complete interacting
massive higher spin Lagrangian for $\Phi$, as there are well-known difficulties
that must be overcome to avoid the presence of ghosts and other unphysical
behavior.  However, for our purposes, we will not require the complete
Lagrangian, as in the end we only need to extract the on-shell behavior of
$\Phi$, and may arrange by hand to place it in an appropriate physical helicity
state.  Essentially, we never need to worry directly about the {\it propagation}
of $\Phi$, but are only interested in its static quantities such as spin and
mass.  Thus, to make $\Phi$ correspond to a leading Regge field, we simply
take it to be transverse and trace-free by assumption.

The effective Lagrangian that we wish to reconstruct is of the form
$\mathcal L=\mathcal L_0+\mathcal L_\Phi$ where
\begin{equation}
e^{-1}\mathcal L_0=R-\ft1{4\cdot5!}F_{(5)}^2-\cdots
\label{eq:iiblag}
\end{equation}
is the usual IIB supergravity Lagrangian (where $5$-form self-duality must
be imposed by hand after obtaining the equations of motion).  For
$\mathcal L_\Phi$, we only consider its coupling to the $5$-form and gravity.
Thus we may write down a diffeomorphism invariant Lagrangian for $\Phi$ of the
form
\bea
e^{-1}\mathcal{L}_{\Phi}&=&
\frac{1}{4\kappa_{10}^2}\Bigl[\ft12
\Phi_{\mu_1\cdots\mu_{2n}}\nabla_\lambda\nabla^\lambda\Phi^{\mu_1\cdots\mu_{2n}}
-\ft12M_\Phi^2\Phi_{\mu_1\cdots \mu_{2n}}\Phi^{\mu_1\cdots \mu_{2n}} \nonumber\\
&&\qquad+\ft12\alpha(n)\Phi_{\mu_1\cdots\mu_{2n}}R^{\mu_1 \nu_1 \mu_2 \nu_2}
\Phi_{\nu_1\nu_2}{}^{\mu_3\cdots\mu_{2n}}
+\ft12\beta(n)\Phi_{\mu_1\cdots\mu_{2n}}R^{\mu_1 \nu_1}
\Phi_{\nu_1}{}^{\mu_2\cdots\mu_{2n}}\nonumber\\
&&\qquad+\ft12\gamma(n) \Phi_{\mu_1\cdots\mu_{2n}}R
\Phi^{\mu_1\cdots\mu_{2n}}\nonumber\\
&&\qquad+\ft14\delta(n)\Phi_{\mu_1\cdots\mu_{2n}}
F^{\mu_1 \nu_1 \alpha_3\cdots \alpha_5}
F^{\mu_2 \nu_2}{}_{\alpha_3\cdots \alpha_5}
\Phi_{\nu_1\nu_2}{}^{\mu_3\cdots\mu_{2n}}\nonumber\\
&&\qquad+\ft14\epsilon(n)\Phi_{\mu_1\cdots\mu_{2n}}
F^{\mu_1 \alpha_2\cdots \alpha_5}F^{\nu_1}{}_{\alpha_2\cdots \alpha_5}
\Phi_{\nu_1}{}^{\mu_2\cdots\mu_{2n}}\nonumber\\
&&\qquad+\ft14\theta(n)\Phi_{\mu_1\cdots\mu_{2n}}
F^{\alpha_1\cdots \alpha_5}F_{\alpha_1\cdots \alpha_5}
\Phi^{\mu_1\cdots\mu_{2n}}+\cdots\Bigr].
\label{GenLag}
\eea
Once again, we do not claim this to be a complete description for a
massive higher spin field $\Phi$.  All that we require is that it reproduces
the equation of motion
\begin{equation}
[\nabla_\lambda\nabla^\lambda-M_{\Phi}^2+\cdots]\Phi_{\mu_1\cdots\mu_{2n}}=0,
\end{equation}
valid when we impose transverse-tracelessness on $\Phi$.  In particular, the
equation of motion is always unambiguous, even in a curved background.

Our goal is to extract the coefficients $\alpha(n),\ldots,\theta(n)$ by
comparison with the string $S$-matrix.  To do so, we make use of the
$\Phi$-$h_{\mu\nu}$-$\Phi$ three point scattering amplitude as well the
$\Phi$-$F_{(5)}$-$F_{(5)}$-$\Phi$ four point scattering amplitude.
At the three-point function level, we will see no contributions that will
allow us to determine the coupling constants $\beta(n)$ and $\gamma(n)$
as they vanish on-shell and do not contribute to the scattering process.
Furthermore, the $\theta(n)$ term (which is included because we only
impose self-duality on $F_{(5)}$ {\it after} obtaining the equations of
motion) will also vanish once the on-shell conditions are imposed.

In any case, after obtaining the effective Lagrangian (\ref{GenLag}), we will
only use it to study the propagation of massive modes on a background
that solves the classical equations of motion for the massless fields.
There, whether in AdS$_5\times S^5$ or the Minkowski background, both
the $\theta(n)$ and $\gamma(n)$ terms are unimportant, and hence they
will be omitted in the remainder of this work.
The $\beta(n)$ term may at first appear troubling since $R_{\mu\nu}$ is
non-vanishing on the AdS$_5\times S^5$ background, but in fact it
also conveniently cancels out on the full equations of motion.
To see this, we first note that the $\beta(n)$ term does not contribute to
a three point scattering process because $R_{\mu \nu}\propto \partial^2
h_{\mu \nu}$ when expanded about flat space, and so the external
graviton leg, which is on shell, gives $\partial^2 h_{\mu \nu}=0$.
However, this three point function will contribute to the four point
$S$-matrix $\Phi$-$F_{(5)}$-$F_{(5)}$-$\Phi$ through a single graviton
exchange diagram obtained by combining the
$\Phi$-$\partial^2h_{\mu\nu}$-$\Phi$ and $F_{(5)}$-$h_{\mu\nu}$-$F_{(5)}$
three point functions with a graviton propagator.  The propagator and
$\partial^2$ from the graviton vertex cancel, leading to a non-pole
contribution to the four-point function which must be canceled by the
$\epsilon(n)$ contact term in (\ref{GenLag}).  The result is that
$\epsilon(n)$ must be related to $\beta(n)$ in just the combination
\be
\ft12\beta(n)\Phi^{\mu_1\mu_2\cdots\mu_{2n}}\Phi^{\nu_1}{}_{\mu_2\cdots\mu_{2n}}
\left(R_{\mu_1 \nu_1}-\frac1{4\cdot4!}F_{\mu_1 \alpha_2 \cdots \alpha_5}
F_{\nu_1}{}^{\alpha_2 \cdots \alpha_5}\right),
\ee
which vanishes on the Einstein equation of motion.  This renders the term
unimportant for our purposes.

Given the effective Lagrangian (\ref{GenLag}), we may then consider its effect
on the propagation of $\Phi$ in the Freund-Rubin background.  At this
quadratic order in $\Phi$, interactions with the background curvature simply
shift the effective mass of $\Phi$ away from its flat ten-dimensional value.
In this manner, we may explore the short-string correction to the flat-space
Regge behavior $m\sim \sqrt S$.  Interestingly, this shift will depend on the
orientation of the ten-dimensional spin of the string.  For example, in a maximally symmetric space, $R_{\mu \nu \rho \sigma}=\pm L^{-2}
\left(g_{\mu \rho} g_{\nu \sigma} - g_{\mu \sigma}g_{\nu \rho}\right)$ where
$L$ is the radius.  Inserting this in (\ref{GenLag}), and assuming an
AdS$_5\times S^5$ background, different components of $\Phi$ will get opposite
sign contributions to their effective mass terms, depending on whether the
spin indices are in the AdS$_5$ or $S^5$ directions.

One may of course worry about the validity of the above Lagrangian, and think
that other higher point interactions are as important as the ones we have
calculated.  We will address this issue below.  However, the end result is that
the regime of validity for our calculations is $n\alpha'\ll L^2$ where
$\alpha'M^2_{\phi}=4(n-1)$ is the flat space mass squared of the $\Phi$
particle, and $L$ is the radius of AdS$_5$.  In this way, the perturbative
results are valid in the large radius limit, where the string theory should
be well approximated by the flat space case.

%%%%%%%%%%%%%%%%%%%%%%%%%%%%%%
\section{The $\Phi$-$h_{\mu\nu}$-$\Phi$ three-point function}

We begin with the simpler of the two scattering computations, namely the
$\Phi$-$h_{\mu\nu}$-$\Phi$ three-point function.  This interaction was
previously worked out in \cite{Giannakis:1998wi} when obtaining the
corrections to the gravitational quadrupole moment of massive string states.
Hence we will be suitably brief, although we do present some of our string
normalization conventions with care.

\subsection{The string SCFT calculation}

We use the standard procedure of computing closed string scattering
amplitudes by first working with one side of the string at a time, and then
``doubling'' the calculation to obtain the closed string result.  Working
in a convenient picture, we use the vertex operators for one side of the
string given by
\bea
V^{\{\mu_n\}}_{-1} &=& \left(i\sqrt{\frac{2}{\alpha'}}
\right)^{(n-1)}\zeta_{\mu_1\cdots \mu_n}e^{-\phi}\psi^{\mu_1}\dx^{\mu_2}\cdots \dx^{\mu_n} e^{ik\cdot X},  \label{eq:vmassive}\\
V^{\{\mu_1\}}_{0} &=&
\xi_{\mu} i\left(\sqrt{\frac{2}{\alpha'}} \dx^{\mu_1}+i\sqrt{\frac{\alpha'}2}\psi^{\mu_1}k\cdot\psi\right)e^{ik\cdot X},
\eea
where we use $\alpha'/2$ rather than $2\alpha'$ because we will only be using our results for closed string amplitudes.
The polarization tensors are symmetric, transverse and trace free:
\bea
\zeta_{\mu_1\cdots\mu_i\cdots\mu_j\cdots\mu_n}
&=&\zeta_{\mu_1\cdots\mu_j\cdots\mu_i\cdots \mu_n},\nonumber \\
k^{\mu_i}\zeta_{\mu_1\cdots\mu_i\cdots \mu_n}&=&0,\nn\\
\eta^{\mu_i\mu_j}\zeta_{\mu_1\cdots\mu_i\cdots\mu_j\cdots \mu_n}&=&0,
\eea
and satisfy the closed string mass shell condition
\be
-k^2=M^2=\frac{2(n-1)}{\alpha'/2}.
\ee

On a single side of the closed string, the three-point function is of
the form
\be
\mathcal A=g_c^{\vphantom\prime}g_c'g_c'\langle
V^{\{\alpha_1\}}_0(z_1) V^{\{\mu_n\}}_{-1}(z_2)V^{\{\nu_n\}}_{-1}(z_3) \rangle,
\label{eq:3ptfn}
\ee
and in general contains terms up to high order in $\alpha'$.  However,
we are only interested in the leading order in $\alpha'$.  In this case,
one must contract the $\dx$ operators together and avoid contractions
with exponentials.  The reason it is consistent to ignore the higher
order terms is because contractions yielding such terms will bring down
factors of momentum from one massive and one massless vertex operator.
Since this is all performed on shell, momentum conservation and transversality
always allows us to replace the massive momentum with a light one
({\it i.e.}\ that of the graviton).  As a result, $\sqrt{\alpha'}k$ where
$k$ is the graviton momentum serves as the expansion parameter; for curvatures
on the scale of $1/L$, this is equivalent to expanding in $\lambda^{-1/4}$.
Dropping the higher order terms then corresponds to scattering in the
low energy limit.

Paying close attention to the various factors that arise in the calculation,
we recall that closed string scattering is conventionally normalized by an
overall factor of
\be
\frac{8\pi}{g_c^2\alpha'}=
\frac{4\pi}{g_c^2}\sqrt{\frac{2}{\alpha'}}\sqrt{\frac{2}{\alpha'}}.
\ee
We find it convenient to absorb one factor of $\sqrt{2/\alpha'}$ into each
side of the calculation, leaving only the more geometric normalization
$4\pi/g_c^2$ as an overall factor.  In this case, the one-sided three point
amplitude (\ref{eq:3ptfn}) breaks up into two pieces, $\mathcal A
=\mathcal A_1+\mathcal A_2$, each of which factors
into bosonic, fermionic and ghost contributions
\bea
\mathcal A_1&=&i
\left(\frac{2}{\alpha'}\right)^n\langle\dx^{\alpha_1}e^{ik_1\cdot X}(z_1)\;
\dx^{\mu_2}\cdots\dx^{\mu_n}e^{ik_2\cdot X}(z_2)\;
\dx^{\nu_2}\cdots\dx^{\nu_n}e^{ik_3\cdot X}(z_3)\rangle\nonumber \\
&&\qquad\qquad\times \left<\psi^{\mu_1}(z_2)\psi^{\nu_1}(z_3)\right>
\langle c(z_1)c(z_2)c(z_3)\rangle\langle e^{-\phi}(z_2)e^{-\phi}(z_3)\rangle,
\nonumber \\
\mathcal A_2&=&-
\left(\frac{2}{\alpha'}\right)^{n-1}\!\!\langle e^{ik_1\cdot X}(z_1)\;
\dx^{\mu_2}\cdots\dx^{\mu_n}e^{ik_2\cdot X}(z_2)\;
\dx^{\nu_2}\cdots\dx^{\nu_n}e^{ik_3\cdot X}(z_3)\rangle\nonumber \\
&&\qquad\qquad\times\left<\psi^{\alpha_1}k_1\cdot \psi(z_1)\psi^{\mu_1}(z_2)
\psi^{\nu_1}(z_3)\right>
\left<c(z_1)c(z_2)c(z_3)\right>\langle e^{-\phi}(z_2)e^{-\phi}(z_3)\rangle.
\quad\label{eq:a1a2}
\eea
In the $\mathcal A_2$ term, one may contract all of the $\dx$ operators with
each other, and so get the leading $\alpha'$ behavior relatively easily.  For
the first term, on the other hand, one must contract at least one of the $\dx$
operators with an exponential, thus bringing down a momentum factor.  Thus to
leading order in $\alpha'$, the resulting amplitude is linear in momenta,
and ends up having the form
\be
\mathcal A=\ft12(n-1)!\left(k_{23}^{\alpha_1}\eta^{\mu_1 \nu_1}
+nk_{12}^{\nu_1}\eta^{\alpha_1 \mu_1}+nk_{31}^{\mu_1}\eta^{\alpha_1 \nu_1}
\right)\prod_{i=2}^n\eta^{\mu_i \nu_i},
\ee
where we have introduced the notation $k_{ij}=k_i-k_j$, and additionally
dropped the overall momentum conserving delta function.  Note that, although
we have stripped the polarization tensors, reserving these for the closed
string amplitude, the above indices must be taken fully symmetric,
$(\mu_1\cdots\mu_n)$ and $(\nu_1\cdots\nu_n)$.  This expression
reduces to the well known trilinear open string gauge boson coupling for
the case of $n=1$.  In general, there are of course higher order in
momentum terms of the form $\alpha'k^2$ in the three-point amplitudes
(\ref{eq:a1a2}).  However, any heavy momenta can be traded off for a light
one through momentum conservation and transversality of the polarization,
leading only to contributions of higher order in $\alpha'/L^2$, which we
may consistently drop.

To convert the above into a closed string amplitude, we ``square'' the
open string result and include a fully-symmetric transverse traceless
polarization tensor with $2n$ indices ($n$ on each side).
Including all factors of $g_c$, as well as the overall normalization,
yields the three-point amplitude with the graviton
\bea
\mathcal A^{\Phi\hbox{-}h_{\mu\nu}\hbox{-}\Phi}
&=&i(2\pi)^{10}\delta^{10}\left({\Sigma k_i}\right)
\left(\frac{1}{2\cdot2\cdot(2 \pi)^2g_c^2}\right)\nn \\
&&\times\left(4\pi g_c\xi^1_{\alpha_{1} \alpha_{2}}\right)
\left(4\pi g_c'(n-1)!\zeta^2_{\mu_1 \cdots \mu_{2n}}\right)
\left(4\pi g_c'(n-1)!\zeta^3_{\nu_1 \cdots \nu_{2n}}\right)
\prod_{i=3}^{2n}\eta^{\mu_i \nu_i}\nonumber\\
&&\times\ft12\left(k_{23}^{\alpha_1}\eta^{\mu_1 \nu_1}+nk_{12}^{\nu_1}\eta^{\alpha_1 \mu_1}+nk_{31}^{\mu_1}\eta^{\alpha_1 \nu_1}\right)
\ft12\left(k_{23}^{\alpha_2}\eta^{\mu_2 \nu_2}+nk_{12}^{\nu_2}\eta^{\alpha_2 \mu_2}+nk_{31}^{\mu_2}\eta^{\alpha_2 \nu_2}\right),\nonumber\\
\label{eq:clo3pt}
\eea
where the corresponding polarizations and momenta are labeled by $1$, $2$
and $3$ for the graviton, first, and second heavy field, respectively.
We have also grouped in canonical factors of $g_c$ and $\pi$ along
with the polarization tensors; this association corresponds to the canonical
normalization used in the effective Lagrangian, below.  We also note here
that the factors of $(n-1)!$ are natural because one may determine $g_c'$ in
terms of $g_c$ from unitarity and the operator/state isomorphism, giving
the relation $g_c'=g_c/(n-1)!$.
%%%%%%%%%%%%%%%%%%%%%%%%%%%%%%%%%%%%%%%%%%%%%%%%%%%%%%%%%%%%%%%%%%%%%%%%%%%%%%

\subsection{The effective Lagrangian}

We now seek the structure of the effective Lagrangian which reproduces the
three-point scattering amplitude, (\ref{eq:clo3pt}).  Expanding the last
line of (\ref{eq:clo3pt}), while making use of the symmetries and transversality
of the polarization tensors, we find
\bea
&&\kern-2em\ft12\left(k_{23}^{\alpha_1}\eta^{\mu_1 \nu_1}
+nk_{12}^{\nu_1}\eta^{\alpha_1 \mu_1}+nk_{31}^{\mu_1}
\eta^{\alpha_1 \nu_1}\right)
\ft12\left(k_{23}^{\alpha_2}\eta^{\mu_2 \nu_2}+nk_{12}^{\nu_2}\eta^{\alpha_2 \mu_2}+nk_{31}^{\mu_2}\eta^{\alpha_2 \nu_2}\right) \nonumber \\
&&\cong k^{\alpha_1}_3k^{\alpha_2}_3 \eta^{\mu_1 \nu_1}\eta^{\mu_2 \nu_2}
+2nk_1^{\mu_1}k_3^{\alpha_1}\eta^{\alpha_2 \nu_1}\eta^{\mu_2 \nu_2}
-2nk_1^{\nu_1}k_3^{\alpha_1}\eta^{\alpha_2 \mu_1}\eta^{\mu_2 \nu_2}
-2n k_3^2 \eta^{\mu_1 \alpha_1} \eta^{\nu_1 \alpha_2}\eta^{\mu_2 \nu_2} \nonumber \\
&&\qquad - 2n M^2 \eta^{\mu_1 \alpha_1} \eta^{\nu_1 \alpha_2}\eta^{\mu_2 \nu_2} \nonumber\\
&&\qquad+ n^2k_1^{\sigma} k_1^{\tilde{\sigma}}
\left(\eta^{\nu_1}_{\sigma}\eta^{\mu_1\alpha_1}-\eta^{\mu_1}_{\sigma}\eta^{\nu_1\alpha_1}\right)
\left(\eta^{\nu_2}_{\tilde{\sigma}}\eta^{\mu_2\alpha_2}-\eta^{\mu_2}_{\tilde{\sigma}}\eta^{\nu_2\alpha_2}\right),
\label{scat}
\eea
where we have made use of $k_3^2=-M^2$ to introduce the mass term.
Here, we use $\cong$ to indicate the expressions are equal when the
polarizations are included.  This may then be identified with the terms
in an effective Lagrangian (for $n>1$)
\bea
e^{-1}\mathcal{L}_\Phi&=&
\frac{1}{4\kappa_{10}^2}\Bigl[\ft12\Phi_{\mu_1\cdots \mu_{2n}} \nabla_\lambda
\nabla^\lambda \Phi_{\mu_1\cdots \mu_{2n}}
-\ft12M^2\Phi_{\mu_1\cdots \mu_{2n}}\Phi_{\mu_1\cdots \mu_{2n}}\nonumber\\
&&\qquad+n^2\Phi_{\mu_1\cdots\mu_{2n}}R^{\mu_1 \nu_1 \mu_2 \nu_2}
\Phi_{\nu_1\nu_2}{}^{\mu_3\cdots\mu_{2n}}
+\beta(n)\Phi_{\mu_1\cdots\mu_{2n}}R^{\mu_1 \nu_1}
\Phi_{\nu_1}{}^{\mu_2\cdots\nu_{2n}}+\cdots\Bigr],\quad~
\label{3ptlag}
\eea
where we have identified $\kappa_{10}=2\pi g_c$ as usual.  The term
proportional to the Ricci tensor is undetermined by on-shell scattering,
and is included with an undetermined coefficient $\beta(n)$.  Note that
the three lines on the right-hand side of (\ref{scat}) correspond directly
to the first three terms in the effective Lagrangian.

The $n=1$ case is different because now all three particles are
indistinguishable.  In fact, this is just the 3 graviton scattering,
and is given by the expansion of the Einstein-Hilbert action to third order
\bea
\frac{1}{2\kappa_{10}^2}\int d^{10}x\sqrt{-g} R\Big|_{3^{rd} order}&=&\nn\\
&&\kern-12em-\frac{1}{2\kappa_{10}^2}\int d^{10}x\ft{1}{4}
\left(h^{\mu_1 \mu_2}h^{\nu_1 \nu_2}\partial_{\mu_1}\partial_{\mu_2}h_{\nu_1 \nu_2}+
2\partial_{\nu_2} h^{\mu_1 \mu_2}\partial_{\mu_2} h_{\mu_1 \nu_1} h^{\nu_1 \nu_2}
+h^{\mu_1 \lambda}h_{\lambda}{}^{\mu_2}\partial_{\sigma} \partial^{\sigma} h_{\mu_1 \mu_2}\right),\nn\\
\eea
where we have taken $g_{\mu \nu}= \eta_{\mu \nu} + h_{\mu \nu}$ and furthermore
assumed that the graviton is transverse and traceless.  The last term in the
above action gives no contribution to the on-shell 3-point scattering amplitude.
It is, however, important in determining the contact term extracted from the
four point scattering, as will be explained below.

%%%%%%%%%%%%%%%%%%%%%%%%%%%%%%%%%%%%%%%%%%%%%%%%%%%%%%%%%%%%%%%%%%%%%%%%%%%%%%
\section{The $\Phi$-$F_{(5)}$-$F_{(5)}$-$\Phi$ four-point function}

We now move on to the interactions of the higher spin field $\Phi$ with the
Ramond-Ramond background.  Due to the underlying fermionic nature of the
Ramond-Ramond states, they only enter pairwise in interactions with the
NSNS field $\Phi$.  As a result, the first non-trivial interaction is at
the four-point level.  Here, an important new feature arises in that the
four-point amplitude factorizes on $s$, $t$ and $u$ channel poles, and that
these underlying three-point terms must be removed.  This, along with
field redefinition ambiguities, somewhat complicate the calculation, although
the methods are standard, at least since the work of
\cite{Gross:1986iv,Gross:1986mw}, which computed the $R^4$ corrections from
closed strings.

\subsection{The SCFT calculation}

For the four-point function, we will again make use of the massive vertex
operators  for $\Phi$ in the $-1$ picture given by (\ref{eq:vmassive}).
The Ramond-Ramond five-form vertex is composed of fermion emission vertex
operators on each side of the string.  On a single side, we make use of
both $1/2$ and $-1/2$ pictures.  In this case, the relevant vertex operators
are
\bea
V_{-1/2}&=&u_{\dot{\alpha}}S^{\dot{\alpha}}_{-1/2}e^{ik\cdot X},\nn \\
V_{1/2}&=&\sqrt{\fft12}\sqrt{\frac{2}{\alpha'}}
\left[i\dx^{\mu}+\fft14\frac{\alpha'}{2}k\cdot\psi\psi^{\mu}\right]
u_{\dot{\alpha}}\Gamma_\mu{}^{\dot{\alpha}}{}_{\beta}S^{\beta}_{1/2}
e^{ik\cdot X},
\eea
where the subscripts on $S_q$ and $\psi_q$ specify the ghost charge,
$e^{q\phi}$.  On a single side, the four-point amplitude given by
\be
\mathcal A=g_c g_c g_c' g_c'\langle V_{1/2}(z_1) V^{\{\mu_n\}}_{-1}(z_2)
V^{\{\nu_n\}}_{-1}(z_3) V_{-1/2}(z_4) \rangle.
\ee
We again find it useful to push one factor of $\sqrt{2/{\alpha'}}$ from the
normalization into the calculation for left movers, and again calculate only
to leading order in $\alpha'$.

This expansion proceeds in much the same way
as the last section, noting again that heavy momenta may only be brought down
by $\dx$ operators from the light vertex operators.  The particulars of the
calculation, while straightforward, are somewhat tedious.  One performs the
$\dx$ correlators in much the same way as the previous calculation.  To find
the contribution from the $\psi$ CFT, one may  use the $SO(10)$ current
algebra $\psi^{\mu} \psi^{\nu}\sim J^{\mu \nu}$, applying it to the $\psi$
and $S$ operators, which transform as the $\mathbf{10}$ and $\mathbf{16}$
respectively.  This reduces all $\psi$ CFT amplitudes to the form
$\langle S(z_1)\psi_{-1}(z_2)\psi_{-1}(z_3) S(z_4)\rangle$, which are known
from scattering of massless string states.
Using these results gives the following amplitude for the left-movers:
\bea
\mathcal A&=&\zeta^2_{\mu_1\cdots \mu_n}\zeta^3_{\nu_1\cdots\nu_n}
u^1_{\dot{\alpha}}u^{4\,\beta}
\frac{(n-1)!}{\sqrt{2}}(-1)^n\prod_{i=2}^{n} \eta^{\mu_i \nu_i}\nonumber\\
&&\quad\times(1-z_3)^{(-t/2-1)}(z_3)^{(-s/2+n-2)}
\left[(1-z_3)A_{(k)}^{\mu_1\nu_1}+ z_3 B_{(k_{23})}^{\mu_1\nu_1}
\right]^{\dot{\alpha}}{}_{\beta}\,,
\eea
where we have taken $z_1\rightarrow \infty, z_2=1, z_4\rightarrow 0$,
$k=k_3+k_4$, and again $k_{ij}=k_i-k_j$.  We
will leave the $z_3$ integration until after including the right-movers.
We have defined a unitless set of Mandelstam variables
\be
s=-\frac{\alpha'}{2}\left(k_1+k_2\right)^2,\qquad
t=-\frac{\alpha'}{2}\left(k_2+k_3\right)^2,\qquad
u=-\frac{\alpha'}{2}\left(k_1+k_3\right)^2,
\ee
and have introduced the kinematical coefficients
\bea
A^{\mu_1 \nu_1}_{(k)}&=&-\ft12(k^{\lambda})
\left(\Gamma^{\mu_1}\Gamma_{\lambda}\Gamma^{\nu_1}\right)
-(n-1)\left(\Gamma^{\nu_1}k^{\mu_1}+\Gamma^{\mu_1}k^{\nu_1}\right),
\nonumber\\
B^{\mu_1 \nu_1}_{(k_{23})}&=&\ft12\left(2\Gamma^{\mu_1}k_{23}^{\nu_1}
+2\Gamma^{\nu_1}k_{23}^{\mu_1}-k_{23}\cdot\Gamma\eta^{\mu_1\nu_1}\right)
+(n-1)\left(\Gamma^{\nu_1}k_{23}^{\mu_1}+\Gamma^{\mu_1}k_{23}^{\nu_1}\right),
\eea
to write the scattering amplitude more succinctly.

For doubling the CFT calculation into a closed string amplitude, one must
take a transpose of the Dirac matrices coming from the right movers.  This
is accounted for by switching the order of the $\mu$ and $\nu$ in the above
coefficients.  The NS polarizations are changed over to closed string
polarizations in the same way as the last calculation.  For the Ramond vertex
operators, on the other hand, we switch to closed string polarizations by
replacing spinor bilinears $u\overline{u}$ with $f_{\{\alpha_i\}}
\Gamma_{(5)}^{\{\alpha_i\}}$ to obtain $F_{(5)}$ polarization states.
Doubling the calculation in this way gives the closed string amplitude
\bea
\mathcal A^{\Phi\hbox{-}F\hbox{-}F\hbox{-}\Phi}
&=&i(2\pi)^{10}\delta\left(\Sigma k_i\right)
\frac{(g_c'(n-1)!)^2(4\pi)}{{2  }}
\left(\zeta^2_{\mu_1\cdots \mu_{2n}}\right)
\left(\zeta^3_{\nu_1\cdots\nu_{2n}}\right)
\bigl(\hat{f}^1_{\alpha_1\cdots \alpha_5}\bigr)
\bigl(\hat{f}^4_{\beta_1\cdots \beta_5}\bigr)
\prod_{i=3}^{2n}\eta^{\mu_i\nu_i}  \nn \\
&&\times\int dz_3 d\bar{z}_3 |(1-z_3)|^{2(-t/2-1)}|z_3|^{2(-s/2+n-2)}\nn \\
&&\qquad\mbox{Tr}
\left[
\Gamma^{\alpha_1 \cdots \alpha_5}
\Big((1-z_3)A^{\mu_1 \nu_1} + z_3B^{\mu_1 \nu_1}\Big)
\Gamma^{\beta_1 \cdots \beta_5}
\Big((1-\bar{z}_3)A^{\nu_2 \mu_2} + \bar{z}_3B^{\nu_2 \mu_2}\Big)
\right]\nn \\
&=&i(2\pi)^{10}\delta\left(\Sigma k_i\right)
(g_c'(n-1)!)^2 4\pi^2
\left(\zeta^2_{\mu_1\cdots \mu_{2n}}\right)
\left(\zeta^3_{\nu_1\cdots\nu_{2n}}\right)
\bigl(\hat{f}^1_{\alpha_1\cdots \alpha_5}\bigr)
\bigl(\hat{f}^4_{\beta_1\cdots \beta_5}\bigr)
\prod_{i=3}^{2n} \eta^{\mu_i \nu_i}\nn\\
&&\times
\frac{\Gamma\left(-\frac{t}{2}\right)\Gamma\left(-\frac{s}{2}+n-1\right)\Gamma\left(-\frac{u}{2}+n-1\right)}
{\Gamma\left(1-\left(-\frac{s}{2}+n-1\right)\right)
\Gamma\left(1-\left(-\frac{u}{2}+n-1\right)\right)\Gamma\left(1-\left(-\frac{t}{2}\right)\right)}\nn\\
&&\times\mbox{Tr}\Big[
\Gamma^{\alpha_1 \cdots \alpha_5}
\Big(-\ft{t}{2}A^{\mu_1 \nu_1}_{(k)} + \left(-\ft{s}{2}+n-1\right)
B^{\mu_1 \nu_1}_{(k_{23})}\Big)\nn\\
&&\kern8em\Gamma^{\beta_1 \cdots \beta_5}
\Big(-\ft{t}{2}A^{\nu_2 \mu_2}_{(k)} + \left(-\ft{s}{2}+n-1\right)
B^{\nu_2 \mu_2}_{(k_{23})}\Big)\Big].
\eea
Again, we are concerned with the limit where the massless fields have zero
momentum.  In this limit, the Mandelstam variables go to their minimum
physical values.  This allows us to see the pole structure arising from the
exchange of on shell intermediate particles, and for doing so we take the
convenient variables
\be
a =\frac{s}{2}-(n-1),\qquad
b =\frac{t}{2},\qquad
c =\frac{u}{2}-(n-1),
\ee
which satisfy $a+b+c=0$. In the limit of low energy scattering $\{a,b,c\}\rightarrow 0$.  After some simple manipulation, we that in this limit
the scattering completely factorizes
\bea
\mathcal A^{\Phi\hbox{-}F\hbox{-}F\hbox{-}\Phi}&=&
i(2\pi)^{10}\delta\left(\Sigma k_i\right)
(g_c'(n-1)!)^2 4\pi^2
\left(\zeta^2_{\mu_1\cdots \mu_{2n}}\right)
\left(\zeta^3_{\nu_1\cdots\nu_{2n}}\right)
\bigl(\hat{f}^1_{\alpha_1\cdots \alpha_5}\bigr)
\bigl(\hat{f}^4_{\beta_1\cdots \beta_5}\bigr)
\prod_{i=3}^{2n} \eta^{\mu_i \nu_i}\nn\\
&&\times\mbox{Tr}
\Bigg[
\frac{\Gamma^{\alpha_1 \cdots \alpha_5}A^{\mu_1 \nu_1}_{(k)}
\Gamma^{\beta_1 \cdots \beta_5}A^{\nu_2 \mu_2}_{(k)}}{a}
+\frac{\Gamma^{\alpha_1 \cdots \alpha_5}B^{\mu_1 \nu_1}_{(k_{23})}
\Gamma^{\beta_1 \cdots \beta_5}B^{\nu_2 \mu_2}_{(k_{23})}}{b}\nn \\
&&\kern6em+\frac{\Gamma^{\alpha_1 \cdots \alpha_5}
\bigl(A^{\mu_1 \nu_1}_{(k)}-B^{\mu_1 \nu_1}_{(k_{23})}\bigr)
\Gamma^{\beta_1 \cdots \beta_5}
\bigl(A^{\nu_2 \mu_2}_{(k)}-B^{\nu_2 \mu_2}_{(k_{23})}\bigr)}{c}
\Bigg].
\label{lescat}
\eea
We have written the scattering in this form because
$A^{\mu \nu}_{(k_3+k_4)}-B^{\mu \nu}_{(k_{23})}=A^{\nu \mu}_{(k_2+k_4)}$.
This demonstrates explicitly that the first and third terms map into each
other by $s\leftrightarrow u$ interchange, namely $2\leftrightarrow 3$,
and $\mu\leftrightarrow \nu$, along with $a\leftrightarrow c$.  The
$t$-channel term (with the $b$ denominator) is distinct, however.

In field theory, the tree level four-point scattering amplitude arises
as a sum of terms
\begin{equation}
\lower4em\hbox{\includegraphics[width=0.5\textwidth]{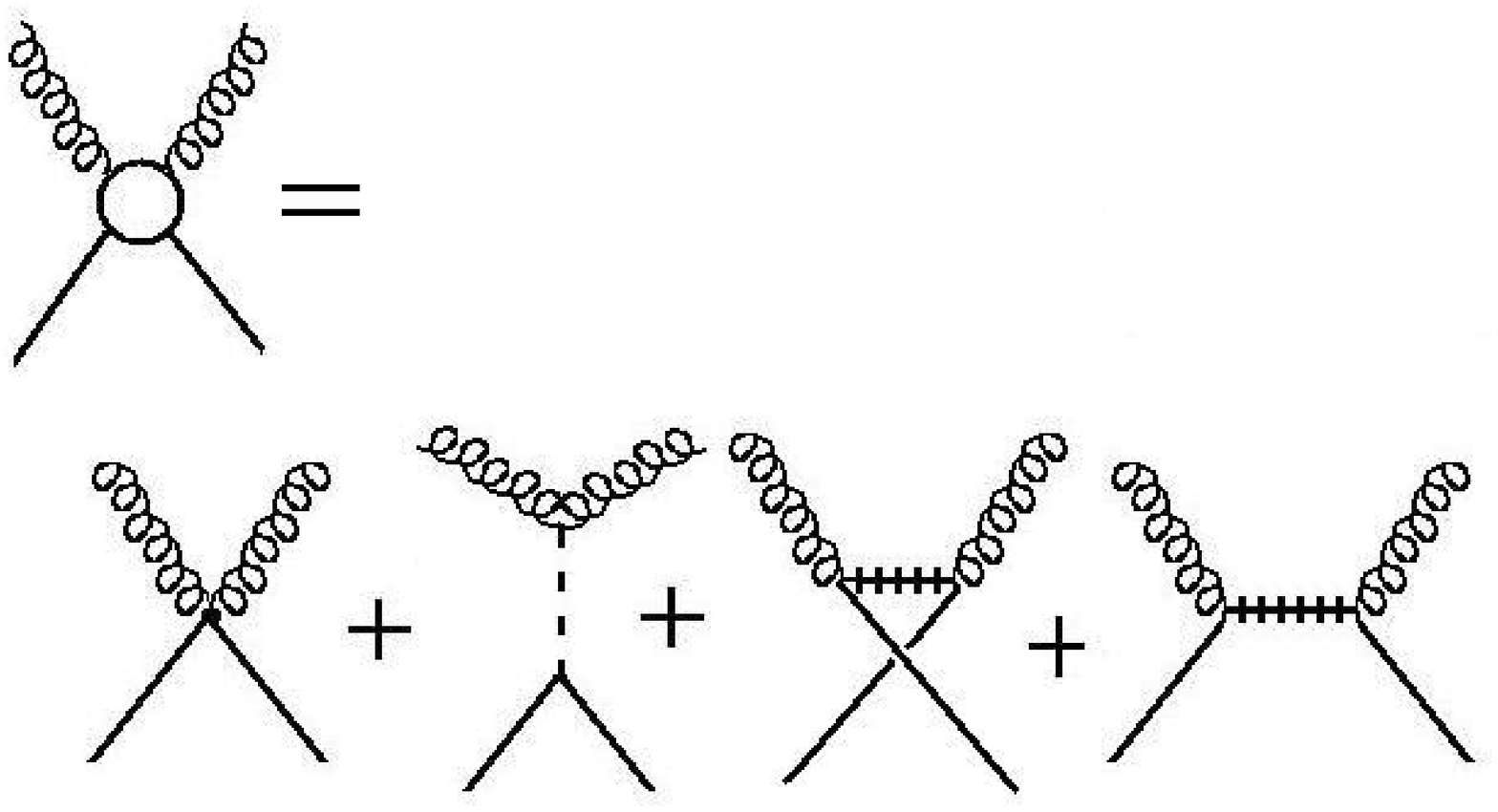}}
\end{equation}
where we use curly lines for $\Phi$, dotted lines for $h_{\mu\nu}$, solid
lines for $F_{(5)}$ and hatched lines to be massive RR fields. (In the
massless case, the curly lines become dotted lines, and the hatches are
removed.)
We wish to extract the contact term which comes directly from the effective
Lagrangian, represented by the term with no intermediate particles.  To do so,
we take the $S$-matrix element calculated in string theory, and subtract
off the ``pole'' terms implied by the three point functions.  This should in
principle completely remove the divergent terms from (\ref{lescat}).

%%%%%%%%%%%%%%%%%%%%%%%%%%%%%%%%%%%%%%%%%%%%%%%%%%%%%%%%%%%%%%
\subsection{Contact terms from the $s$ and $u$ channels}

To properly obtain the four-point contact term from the $S$-matrix, one should
subtract scattering corresponding to the exchange of all intermediate particles
present in the effective Lagrangian.  This will in principle cancel the poles
in the above scattering amplitude, (\ref{lescat}).  Unfortunately, this
proves to be a difficult calculation for the $s$ and $u$ channels,
primarily because the propagator for off-shell higher mass RR
fields are difficult to write down.  Because of this, we use an alternate
method to isolate the contact terms from these poles.

Basically, we assume that the field $\Phi$ does not couple via a
$\Phi F_{(5)}\Box f_m$ term, where $f_m$ denotes a massive RR field, and
$\Box$ indicates the equation of motion for $f_m$.  Given this, the underlying
single particle exchange diagrams in the $s$ and $u$ channel will not lead
to constant (non-residue) terms.  As a result, the $s$ and $u$ channel
contributions to the contact term are fully contained in the constant parts
of the corresponding terms in (\ref{lescat}).  (Furthermore, the residues of
the $1/a$ and $1/c$ poles must necessarily be canceled, but this is what
we do not check.)  For such a constant term, there must be a momentum squared
in the numerator to cancel the denominator.  Such a term occurs in the $s$
channel pole because $A^{\mu \nu}_{(k)}$ contains a $k\cdot \Gamma$ piece.
The $u$ channel piece is identical to the $s$ channel, simply switching
$\mu \leftrightarrow \nu$.

Our procedure to identify the contact terms is therefore as follows.
First, we identify terms in the $s$ channel that are proportional to $k^2/2$
in the numerator, and expect that this $k$ dependence cancels the denominator.
This process introduces a factor of $\alpha'/2$ because $k^2$ has units,
while $a$ does not.  This is the usual $\alpha'$ dependence that
is included in the normalization of the $\hat{f}$ polarization tensors, $\hat{f}=\sqrt{\alpha'/2}{f}$.
Taking this as the contact piece in the $s$
channel, one may include the $u$ channel by adding the $\mu\leftrightarrow\nu$
contribution by hand.  This process gives the contact terms to be
\bea
\mathcal A_{s,u}^{4\rm pt}&=&i(2\pi)^{10}\delta\left(\Sigma k_i\right)
4\pi^2(g_c'(n-1)!)^2
\left(\zeta^2_{\mu_1\cdots \mu_{2n}}\right)
\left(\zeta^3_{\nu_1\cdots\nu_{2n}}\right)
\bigl({f}^1_{\alpha_1\cdots \alpha_5}\bigr)
\bigl({f}^4_{\beta_1\cdots \beta_5}\bigr)
\prod_{i=3}^{2n} \eta^{\mu_i \nu_i} \nn\\
&&\quad\times\mbox{Tr}
\left[\Gamma^{\alpha_1 \cdots \alpha_5}\ft12(k^{\lambda})
\left(\Gamma^{\mu_1}\Gamma_{\lambda}\Gamma^{\nu_1}\right)
\Gamma^{\beta_1 \cdots \beta_5}\ft12(k^{\tilde{\lambda}})
\left(\Gamma^{\nu_2}\Gamma_{\tilde{\lambda}}\Gamma^{\mu_2}\right)
+\mu\leftrightarrow \nu
\right]_{-k^2/2 ~\rm factor}\nn\\
&=&i(2\pi)^{10}\delta\left(\Sigma k_i\right)2\pi^2(g_c'(n-1)!)^2
\left(\zeta^2_{\mu_1\cdots \mu_{2n}}\right)
\left(\zeta^3_{\nu_1\cdots\nu_{2n}}\right)
\bigl({f}^1_{\alpha_1\cdots \alpha_5}\bigr)
\bigl({f}^4_{\beta_1\cdots \beta_5}\bigr)
\prod_{i=3}^{2n} \eta^{\mu_i \nu_i}\nonumber\\
&&\quad\times\mbox{Tr}
\left[\left(\Gamma^{\mu_2}\Gamma^{\alpha_1 \cdots \alpha_5}
\Gamma^{\mu_1}\right)\left(\Gamma^{\nu_1}
\Gamma^{\beta_1 \cdots \beta_5}\Gamma^{\nu_2}\right)
+\mu\leftrightarrow \nu\right].
\eea
We take the above trace, remembering that certain terms vanish because of self
duality of the five form, and that many terms are equivalent because of the
symmetries of the polarization tensors.  Our final result for the scattering
due to the contact piece is given by
\bea
\mathcal{A}_{s,u}^{4\rm pt}&=&i(2\pi)^{10}\delta\left(\Sigma k_i\right)
\frac{1}{2\cdot (2 \pi g_c)^2 \cdot 4\cdot 5!}
\left((4\pi g_c'(n-1)!)\zeta^2_{\mu_1\cdots \mu_{2n}}\right)
\left((4\pi g_c'(n-1)!)\zeta^3_{\nu_1\cdots\nu_{2n}}\right)\nn \\
&&\quad\times
\left(4\pi g_c\cdot\sqrt{32}\cdot 5! {f}^1_{\alpha_1\cdots \alpha_5}\right)
\left(4\pi g_c\cdot\sqrt{32}\cdot 5!{f}^4_{\beta_1\cdots \beta_5}\right)
\prod_{i=3}^{2n} \eta^{\mu_i \nu_i}\prod_{j=3}^5\eta^{\alpha_j\beta_j}\nn\\
&&\quad\times\left[
5\eta^{\mu_1 \alpha_1}\eta^{\nu_1 \beta_1}\eta^{\mu_2 \nu_2}
\eta^{\alpha_2 \beta_2}
+5\eta^{\nu_1 \alpha_1}\eta^{\mu_1 \beta_1}\eta^{\mu_2 \nu_2}
\eta^{\alpha_2 \beta_2}
-40\eta^{\nu_1 \alpha_1}\eta^{\mu_1 \beta_1}
\eta^{\nu_2 \beta_2}\eta^{\mu_2 \alpha_2}\right].\qquad
\label{suchanscat}
\eea
We will take this as the $s$ and $u$ channel contributions to contact terms.
This gives the correct result for the massless case where the subtraction
can be done, the intermediate field being the massless five form.

%%%%%%%%%%%%%%%%%%%%%%%%%%%%%%%%%%%%%%%%%%%%%%%%%%%%%%%%%%%%%%%%%%%
\subsection{Contact term from the $t$ channel}

As mentioned above, the difficulty in addressing the exact $s$ or $u$ channel
subtractions is a result of not knowing massive higher spin RR propagators.
However, in the $t$ channel, the only intermediate particle allowed in the
low energy limit is the graviton.  The dilaton does not enter because of self
duality of the five-form.  The fact that this subtraction can be done leads
to a cancellation of the $\beta(n)$ term in our final answer, as we will show in
the following.

First, the field theoretic quantities needed are the graviton propagator,
the Feynman rule for the $\Phi$-$h_{\mu\nu}$-$\Phi$ scattering, and the
Feynman rule for the $F_{(5)}$-$h_{\mu\nu}$-$F_{(5)}$ scattering.
The graviton propagator is
\bea
\langle h^{\mu \nu}(p)h_{\rho \sigma}(q)\rangle
&\equiv&\Delta^{\mu \nu}_{\rho \sigma}(p,q) \nn \\
&=&-i(2\pi)^{10}\delta^{10}\left(p+q\right)
\frac{4\kappa_{10}^2}{p^2}\fft12
\left(\mathcal{P}^\mu_\rho\mathcal{P}^{\nu}_{\sigma}
+\mathcal{P}^\mu_\rho\mathcal{P}^{\nu}_{\sigma}
-\frac{2}{9}\mathcal{P}^{\mu \nu}\mathcal{P}_{\rho \sigma}\right),
\eea
where we have used
\be
\mathcal{P}^\mu_\rho(p)\equiv \left(\eta^{\mu}_{\rho}-\frac{p^{\mu}p_{\rho}}{p^2}\right),
\ee
so that $\Delta$ satisfies the transverse traceless conditions.  We are
evaluating these between conserved currents, one of which is traceless
$(F_{(5)}^2=0)$, so we are left using only
\be
\langle h^{\mu \nu}(p)h_{\rho \sigma}(q)\rangle
=-i(2\pi)^{10}\delta^{10}\left(p+q\right)\frac{4\kappa_{10}^2}{p^2}\fft12
\left(\eta^\mu_\rho\eta^{\nu}_{\sigma}
+\eta^\mu_\rho\eta^{\nu}_{\sigma}+\cdots\right).
\ee
The Feynman rule for
$\Phi^2_{\mu_i}$-$h_{\alpha_1\alpha_2}^{-(2+3)}$-$\Phi^3_{\nu_i}$
scattering can be read off from (\ref{3ptlag}),
\bea
\Gamma&=&
\frac{i}{4(\kappa_{10})^2}\prod_{i=3}^{2n}\eta^{\mu_i \nu_i}
\Bigl[\beta(n)(k_2+k_3)^2\eta^{\mu_2\nu_2}
\left(\eta^{\mu_1 \alpha_1} \eta^{\nu_1 \alpha_2}\right)\nn\\
&&\quad
+\ft12\left(k_{23}^{\alpha_1}\eta^{\mu_1 \nu_1}
-2nk_{23}^{\nu_1}\eta^{\alpha_1 \mu_1}
-2nk_{23}^{\mu_1}\eta^{\alpha_1 \nu_1}\right)
\ft12\left(k_{23}^{\alpha_2}\eta^{\mu_2 \nu_2}
-2nk_{23}^{\nu_2}\eta^{\alpha_2 \mu_2}
-2nk_{23}^{\mu_2}\eta^{\alpha_2 \nu_2}\right)\Bigr],\nn\\
\eea
where we have replaced all graviton momenta with momenta coming from the
$\Phi$ fields, and have taken advantage of the transversality of all
polarizations and propagators involved.  Note that in transverse traceless
gauge $R_{\mu \nu}=-\fft12 \Box h_{\mu \nu}$, which gives the $\beta(n)$
dependent piece.

We now make the useful definition
\be
T^{\alpha \mu \nu}_{(n)}(k_{23})\equiv
\left(k_{23}^{\alpha}\eta^{\mu \nu}-2nk_{23}^{\nu}\eta^{\alpha \mu}-2nk_{23}^{\mu_2}\eta^{\alpha \nu}\right)
\ee
to more easily express scattering amplitudes and Feynman rules.  One should
note that $B^{\mu \nu}=(1/2)T^{\alpha \mu \nu}\Gamma_\alpha$, which will make
the field theoretic and string theoretic calculations easier to compare.
{}From the $F_{(5)}^2$ term in the IIB Lagrangian (\ref{eq:iiblag}),
one can read off the Feynman rule for
$F^1_{\rho_i}$-$h_{\tilde{\alpha}_1\tilde\alpha_2}^{-(1+4)}$-$F^4_{\sigma_i}$
\be
\Gamma=-i\frac{1}{2\kappa_{10}^2}\frac{10}{4\cdot 5!}
\eta^{\tilde{\alpha}_1 \rho_1}\eta^{\tilde{\alpha}_2 \sigma_1}\prod_{i=2}^5\eta^{\sigma_i \rho_i}.
\ee
which can also be verified by examining the string
$F_{(5)}$-$h_{\mu\nu}$-$F_{(5)}$
three-point function.  Computing the field theoretic scattering, remembering
the factor $1/2!$ because it is second order in perturbation theory, one
finds the $t$ channel pole contribution to the scattering amplitude
\bea
\mathcal A_{\rm sub}&=&-i\left(2\pi\right)^{10}\delta\left(\Sigma k_i\right)
\biggl[\frac{1}{2\kappa_{10}^2 4\cdot5!}\left(\frac{5}{2(k_2+k_3)^2}\right)
\Phi^2_{\mu_1 \mu_2}{}^{\nu_3 \cdots \nu_{2n}}
\Phi^3_{\nu_1 \nu_2 \nu_3 \cdots \nu_{2n}}\nn \\
&&\kern14em\times
T^{\rho_1 \mu_1 \nu_1}T^{\sigma_1 \mu_2 \nu_2}
(F^3_{\rho_1 \rho_2 \cdots \rho_5})
(F^4_{\sigma_1}{}^{\rho_2 \cdots \rho_5})  \nn\\
&&\quad+\frac{\beta(n)}{8\kappa_{10}^2}\frac1{4\cdot4!}2
(\Phi^{2\,\mu_1 \mu_2\cdots\mu_{2n}})
(\Phi^{3\,\nu_1}{}_{\mu_2\cdots\mu_{2n}})
\left(F^1_{\mu_1 \rho_2 \cdots \rho_5}
F^4_{\nu_1}{}^{\rho_2 \cdots \rho_5}+
F^4_{\mu_1 \rho_2 \cdots \rho_5}F^1_{\nu_1}{}^{\rho_2 \cdots\rho_5}
\right)\bigg].\nn\\
\eea
To compare this with the string theory calculation, we take the $t$ channel
($1/b$) term from (\ref{lescat}) and evaluate the trace of Dirac matrices,
again making use of the symmetries of the polarizations, and noting that
certain terms are zero due to self duality of the five-form.  The final
result for the $t$ channel string amplitude is
\bea
\mathcal A_{\rm str}&=&-i\left(2\pi\right)^{10}\delta\left(\Sigma k_i\right)
\bigg[
\frac{1}{2\kappa_{10}^2 4\cdot 5!}\left(\frac{5}{2(k_2+k_3)^2}\right)
\Phi^2_{\mu_1 \mu_2}{}^{\nu_3 \cdots \nu_{2n}}
\Phi^3_{\nu_1 \nu_2 \nu_3 \cdots \nu_{2n}}\nn\\
&&\kern14em\times T^{\rho_1 \mu_1 \nu_1}T^{\sigma_1 \mu_2 \nu_2}
(F^3_{\rho_1 \rho_2 \cdots \rho_5})
(F^4_{\sigma_1}{}^{\rho_2 \cdots \rho_5})\bigg].
\eea
As a result, subtracting the underlying graviton exchange
contribution $\mathcal A_{\rm sub}$ cancels the pole completely, while
introducing a $\beta(n)$ dependent term.  This leads to a contact term
with undetermined coefficient $\beta(n)$
\bea
\mathcal{A}_t^{4\rm pt}&=&i\left(2\pi\right)^{10}\delta\left(\Sigma k_i\right)
\frac{\beta(n)}{8\kappa_{10}^2}\frac1{4\cdot4!}2
(\Phi^{2\,\mu_1 \mu_2\cdots\mu_{2n}})
(\Phi^{3\,\nu_1}{}_{\mu_2\cdots\mu_{2n}})\nn\\
&&\kern6em\times \left(F^1_{\mu_1 \rho_2 \cdots \rho_5}
F^4_{\nu_1}{}^{\rho_2 \cdots \rho_5}+
F^4_{\mu_1 \rho_2 \cdots \rho_5}F^1_{\nu_1}{}^{\rho_2 \cdots \rho_5}\right).
\label{tchanscat}
\eea
We will discuss the total Lagrangian and connection to the massless case in
the next subsection.

%%%%%%%%%%%%%%%%%%%%%%%%%%%%%%%%%%%%%%%%%%%%%%%%%%%
\subsection{Complete effective action}

Reading off the three-point effective action from (\ref{3ptlag}), and
adding in the four-point terms from (\ref{suchanscat}) and (\ref{tchanscat})
gives the following effective Lagrangian
\bea
e^{-1}\mathcal{L}_\Phi&=&
\frac{1}{4\kappa_{10}^2}\biggl[
\ft12 \Phi_{\mu_1\cdots \mu_{2n}} \nabla_{\lambda} \nabla^{\lambda}
\Phi^{\mu_1\cdots \mu_{2n}}
-\ft12M_\Phi^2\Phi_{\mu_1\cdots \mu_{2n}}\Phi^{\mu_1\cdots \mu_{2n}}
\nonumber \\
&&\qquad+n^2\Phi_{\mu_1\cdots\mu_{2n}}R^{\mu_1 \nu_1 \mu_2 \nu_2}
\Phi_{\nu_1\nu_2}{}^{\mu_3\cdots\mu_{2n}} \nn \\
&&\qquad-\frac{1}{4\cdot 5!}20\Phi_{\mu_1\cdots\mu_{2n}}
F^{\mu_1 \nu_1 \alpha_3\cdots \alpha_5}
F^{\mu_2 \nu_2}{}_{\alpha_3\cdots \alpha_5}
\Phi_{\nu_1\nu_2}{}^{\mu_3\cdots\mu_{2n}}\nonumber\\
&&\qquad-\frac{1}{4\cdot5!}5\Phi_{\mu_1\cdots\mu_{2n}}
F^{\mu_1 \alpha_2\cdots \alpha_5}F^{\nu_1}{}_{\alpha_2\cdots \alpha_5}
\Phi_{\nu_1}{}^{\mu_2\cdots\mu_{2n}}\nonumber\\
&&\qquad+\ft12\beta(n)\Phi_{\mu_1\cdots\mu_{2n}}
\left(R^{\mu_1 \nu_1}-\fft1{4\cdot4!}(F^2)^{\mu_1 \nu_1}\right)
\Phi_{\nu_1}{}^{\mu_2\cdots\mu_{2n}}\nonumber \\
&&\qquad+\ft12\gamma(n)\Phi_{\mu_1\cdots\mu_{2n}}R\Phi^{\mu_1\cdots\mu_{2n}}
\nn\\
&&\qquad+\ft14\theta(n)\Phi_{\mu_1\cdots\mu_{2n}}
F^{\alpha_1\cdots \alpha_5}F_{\alpha_1\cdots \alpha_5}
\Phi^{\mu_1\cdots\mu_{2n}}+\cdots\biggr].
\label{FinLag}
\eea
Note that the last three lines are couplings to terms that vanish on shell,
and in particular $\beta(n)$ describes the coupling of $\Phi$ to the Einstein
equation.  One may think of this as generalizing an on shell condition,
linearized about flat space, $\Box h_{\mu \nu}=0$, to the full non-linear
equations $R^{\mu_1 \nu_1}-\frac1{4\cdot4!}(F^2)^{\mu_1 \nu_1}=0$
describing a curved background.

We now check that the above Lagrangian indeed reproduces the standard IIB
supergravity action for the massless modes.  First, remember that in the
massless case that the kinetic $\Phi$ terms and the Riemann coupling are
instead identified with on shell parts of $R$ expanded to third order.
Next, for the off shell part, note that the combinatoric factor of $2!$
arising from assigning particle 2 or 3 to each of the $\Phi$ fields in
$\Phi^{\mu \lambda_2\cdots\lambda_{2n}}\Phi_{\nu \lambda_2\cdots\lambda_{2n}}
R^{\nu}_\mu$ is the same $2!$ for that of assigning 2 or 3 to
$-h^{\mu \lambda}h_{\lambda}^{\nu}\Box h_{\mu \nu}$.  This is because 2 or 3
may be assigned to the first two $h$ factors, but not the third because the
external legs are on shell.  This gives that the massless case is an analog
of $\beta(1)=2$.  Plugging this in, one finds that the scattering Lagrangian
becomes
\bea
\mathcal{L}_{hhh}+\mathcal{L}_{hhFF}&=&
\frac{1}{4\kappa_{10}^2}\sqrt{-g}\biggl[-\ft12h^{\mu_0 \nu_0}h^{\mu_1 \nu_1}
\partial_{\mu_0} \partial_{\nu_0} h_{\mu_1\nu_1}
-h^{\mu_1}_{\nu_1}h_{\mu_1 \nu_2, \lambda}h^{\lambda \nu_2 ,\nu_1}\nn\\
&&\qquad+h^{\mu_1\mu_2}h^{\nu_1}_{\mu_{2}}\left(-\fft12\Box h_{\mu_1 \nu_1}-\frac1{4\cdot4!}(F^2)_{\mu_1 \nu_1}\right)
\nonumber \\
&&\qquad-\frac{1}{8\cdot5!}40 h^{\mu_1\mu_{2}}h^{\nu_1\nu_{2}}
\left(F_{\mu_1 \nu_1 \alpha_3\cdots \alpha_5}
F_{\mu_2 \nu_2}{}^{\alpha_3\cdots \alpha_5}\right)
\nonumber \\
&&\qquad-\frac{1}{8\cdot5!}10 h^{\mu_1\mu_{2}}h^{\nu_1}_{\mu_{2}}
\left(F_{\mu_1\alpha_2\cdots\alpha_5}F_{\nu_1}{}^{\alpha_2\cdots\alpha_5}
\right)\biggr]\nn\\
&=&\frac{1}{2\kappa_{10}^2}\left[(\sqrt{-g})R|_{hhh}
-\big(\sqrt{-g}\frac{1}{4\cdot5!} F^2\big)|_{hhFF}
\right]_{\mbox{$F=*F$ imposed}}.
\eea
The $\beta(1)=2$ term contributes to one of the $F_{(5)}^2$ terms in just the
way needed to give the combinatoric factors of $\binom51$ and $\binom52$,
understood as expanding either one of the inverse metrics in $F_{(5)}^2$
to second order, or two of them to first order.

The unknowns $\beta(n)$, $\gamma(n)$ and $\theta(n)$ all multiply terms
which vanish on-shell (and which hence are undetermined from the $S$-matrix
computation).  Of course, taking (\ref{FinLag}) and evaluating $R$ and
$F_{(5)}$ as a background solution, we find all unknown coefficients have
dropped out.  Such a statement may require more care for other background
form fields.  For example, without using equations of motion,
$\sqrt{-g}\Phi^2 R$ zero for a $\Phi \Phi h$ scattering, the first order
expansion of $\sqrt{-g}R$ vanishing in transverse traceless gauge.  This term
is zero, therefore, for a different reason than $\sqrt{-g}(\Phi^2)^{\mu\nu}
R_{\mu \nu}$ is zero, the latter vanishing {\it on shell}.  As a result, one
cannot relate the terms $\sqrt{-g}\Phi^2 R$ and $\sqrt{-g}F_{(3)}^2 \Phi^2$
as one may relate $\sqrt{-g}(\Phi^2)^{\mu\nu}  R_{\mu \nu}$ and
$\sqrt{-g}(\Phi^2)^{\mu\nu} (F_{(3)}^2)_{\mu \nu}$, although it is tempting
to think that these also come in the combination dictated by the
trace of the Einstein equations.  To find these terms, one would use a
$\Phi \Phi hh$ scattering, which will not vanish in transverse traceless
gauge.  As a further consideration, one must also consider exchange of
dilatons in the case of the three form.  We were able to ignore this term
because of the self duality condition on $F_{(5)}$.

%%%%%%%%%%%%%%%%%%%%%%%%%%%%%%%%%%%%%%%%%%%%%%%%%%%%%%%%%%%%%%%%%
\subsection{Regime of Validity}

We now are prepared to discuss the regime of validity for which the above
Lagrangian terms are valid.  First, let us note that we are {\it restricting}
our discussion to terms that are quadratic in $\Phi$, and coupled to fields
with massless quanta.  In this sense, we are focusing on the terms in the
Lagrangian that will change the ``free'' equations of motion in a background.
Also, we will consider only sphere (tree) level amplitudes, and only
backgrounds with metric and five form.

We have used $k_0\sqrt{\alpha'}$ (light momenta) as expansion parameters, and
we will show that we also need to take $n\alpha' \ll L^2$.  We use $L$ to
denote the smallest length scale associated with the space, in the case
of AdS$_5 \times S^5$, this is simply the radius of either space.  Expansions
in $\alpha'$ usually fail because when one considers higher derivatives of
massive fields one finds large factors, as these momenta are inherently larger
than 1 in $\alpha'$ units.  We consider higher order interactions
schematically of the kind $R^i F^{2j} \Phi \Phi$.  Terms with the Ricci scalar
are zero for our purposes, and terms with the Ricci tensor will couple to
the equation of motion as before.  Therefore, we only consider the Riemann
tensor in for $R$.  The above coupling can be determined by a
$h_{(0,0)}^i (F_{(+\fft12,+\fft12)} F_{(-\fft12,-\fft12)})^j
\Phi_{(-1,-1)} \Phi_{(-1,-1)}$ scattering where we keep the  $\Phi$ terms in
their canonical ghost picture.  One might think that the extra $\dx$ operators
on the $\Phi$ vertex operators will cause most of the problem.  However, these
may only bring down momenta from the other $\Phi$ vertex operator, which can
then be replaced by massless momenta thanks to the transversality of the
polarization of $\Phi$.  The $\dx$ operators from massless vertex operators
will lead to heavy derivatives, which might cause a problem.  The maximum
number of heavy derivatives is $2(i+j)$, simply from examining the structure
of the vertex operators (count one $\dx$ and one $\dbx$ per vertex operator
with a ghost picture $+1$ of its canonical value).  Therefore, in the worst
case, one may find $R^i F^{2j} \partial^{2(i+j)}\Phi \Phi$ type interactions.
Taking the center of mass energy to be small is the same as putting the
$\Phi$ particle close to being at rest, so that the derivatives essentially
give $\sqrt{M_\Phi^2}\sim \sqrt{n\alpha'}$.  Also, $R$ and $F^2$ are both of
the order $1/L^2$.  This gives that the overall order of the term is
$(n\alpha'/L^2)^{(i+j)}$.  Thus, to take this term as being small, we must
have $n \alpha' \ll L^2$ as advertised.  One may think of this as quantizing
the string in the vicinity of its classical geodesic.  In the massive case we
are considering, the geodesic is simply a particle sitting at a point in a
smooth manifold, and so the space near the geodesic is flat space.

%%%%%%%%%%%%%%%%%%%%%%%%%%%%%%%%%%%%%%%%%%%%%%%%%%%%%%%%%%%%%%%%%%%%%%%%%%%%%%%%
\section{AdS$_5 \times S^5$ and spinning strings}

Now that we have constructed the effective Lagrangian (\ref{FinLag})
up to first non-trivial order in $\alpha'/L^2$, we may finally investigate
its consequences for spinning strings in an AdS$_5\times S^5$ background.
For a direct product space, we now switch notations so that $M,N,\ldots$
denote ten-dimensional indices, while $\mu,\nu\,\ldots$ and $m,n,\ldots$
denote AdS$_5$ and $S^5$ indices, respectively.  The AdS$_5\times S^5$
background is then a solution of the Freund-Rubin ansatz
\begin{eqnarray}
&&R_{\mu\nu}=-\fft4{L^2}g_{\mu\nu},\kern3.2em R_{mn}=\fft4{L^2}g_{mn},\nn\\
&&F_{\mu\nu\rho\lambda\sigma}=\fft4L\epsilon_{\mu\nu\rho\lambda\sigma},
\qquad F_{mnpqr}=\fft4L\epsilon_{mnpqr},
\end{eqnarray}
solving the equations of motion
\begin{equation}
R_{MN}=\fft1{4\cdot4!}F_{MPQRS}F_N{}^{PQRS},\qquad F=*F,\qquad dF=0.
\end{equation}
For the maximally symmetric AdS$_5\times S^5$ solution, we may furthermore
write
\be
R_{M N I J}=\frac1{4\cdot4!}F_{MNPQR}F_{IJ}{}^{PQR},
\ee
where
\begin{equation}
R_{\mu\nu\rho\sigma}=-\fft1{L^2}(g_{\mu\rho}g_{\nu\sigma}-g_{\mu\sigma}
g_{\nu\rho}),\qquad
R_{mnpq}=\fft1{L^2}(g_{mp}g_{nq}-g_{mq}g_{np}).
\label{eq:riemanns}
\end{equation}
This allows us to replace the five-form terms by curvature terms when
evaluating (\ref{FinLag}) on the AdS$_5\times S^5$ background
\bea
e^{-1}\mathcal{L}_\Phi&=&
\frac{1}{4\kappa_{10}^2}\Bigl[
\ft12\Phi_{M_1\cdots M_{2n}} \nabla_{L}\nabla^L \Phi^{M_1\cdots M_{2n}}
-\ft12M_\Phi^2\Phi_{M_1\cdots M_{2n}}\Phi^{M_1\cdots M_{2n}} \nonumber \\
&&\qquad+(n^2-4) \Phi_{M_1M_2M_3\cdots M_{2n}}R^{M_1N_1M_2N_2}
\Phi_{N_1N_2}{}^{M_3\cdots M_{2n}} \nn \\
&&\qquad-\Phi_{M_1M_2\cdots M_{2n}}R^{M_1N_1}\Phi_{N_1}{}^{M_2 \cdots M_{2n}}
+\cdots\Bigr].
\eea
Using (\ref{eq:riemanns}), as well as (ten-dimensional) tracelessness of
$\Phi$, results in a relatively simply effective Lagrangian
\bea
e^{-1}\mathcal L_\Phi&=&\frac{1}{4\kappa_{10}^2}\Bigl[
\ft12\Phi_{M_1\cdots M_{2n}} \nabla_L\nabla^L \Phi^{M_1\cdots M_{2n}}
-\ft12M_\Phi^2\Phi_{M_1\cdots M_{2n}}\Phi^{M_1\cdots M_{2n}} \nonumber \\
&&\qquad+\frac{n^2}{L^2}\left(
\Phi_{\mu_1 \mu_2 M_3 \cdots M_{2n}}\Phi^{\mu_1\mu_2 M_3\cdots M_{2n}}
-\Phi_{m_1 m_2 M_3 \cdots M_{2n}}\Phi^{m_1m_2M_3\cdots M_{2n}}\right)
+\cdots\Bigr].\qquad
\label{eq:efflag1}
\eea
Note that the background curvature terms on AdS$_5$ and $S^5$ contribute
equally, but with opposite signs.

A symmetric tensor $\Phi_{M_1\cdots M_{2n}}$ in ten dimensions decomposes
into a direct sum of symmetric tensors on AdS$_5$, transforming in
symmetric representations of the $R$-symmetry group $SO(6)$.  Focusing on
a single term in this decomposition, we consider in general a field $\Phi$
with $S$ indices along AdS and $K$ indices along the sphere, and of course
the condition $S + K=2n$.  The fields are symmetrized, and so we may only
choose the number of AdS$_5$ versus $S^5$ indices, and not the order in which
they come.  We consider the field as $\Phi_{\{M_j\}}=(1/(2n)!)
\sum_i \Phi_{Mi}$, where the $M_j$ are given as either AdS$_5$ or $S^5$
indices, and the right hand side gives the full symmetrized field. In this
expression, there will be a given number of times that the AdS$_5$ indices
appear in the first two entries, and a given number of times that the $S^5$
indices appear in the first two entries.  This breakdown is given by a
relatively elementary combinatorics problem.  Of the total $(2n)!$ terms in
the symmetrized sum, the first two entries break down as:
$(2n-2)!S(S-1)$ of them are both AdS$_5$; $(2n-2)!2SK$ entries come as
mixed; and $(2n-2)!K(K-1)$ of them are both $S^5$.  It is easy to check
that $S(S-1)+K(K-1)+2SK=2n(2n-1)$.
Therefore, for a particular mixed index component, we have
\bea
g^{\mu_1 \nu_1}g^{\mu_2 \nu_2} \Phi_{\mu_1 \mu_2 \cdots}
\Phi_{\nu_1 \nu_2 \cdots}
&=&g^{\mu_1 N_1}g^{\mu_2 N_2} \Phi_{\mu_1 \mu_2 \cdots} \Phi_{N_1 N_2 \cdots } \nonumber\\
&=&\frac{S(S-1)}{2n(2n-1)}g^{M_1 N_1}g^{M_2 N_2} \Phi_{M_1 M_2 \cdots}
\Phi_{N_1 N_2 \cdots },
\eea
where in the first step we may generalize the indices because the metric has
no mixed components.  We do this for the sphere metrics as well, replacing $S$
by $K=2n-S$ in the above.  A second way to arrive at the same answer is by
expanding all metrics as $g^{MN}=g^{\mu \nu} + g^{m n}$ and keeping track of
combinatorial terms.  In either case, the Lagrangian (\ref{eq:efflag1})
becomes
\bea
e^{-1}\mathcal L_\Phi&=&\frac{1}{4\kappa_{10}^2}\sum_{S=0}^{2n}
\biggl[\ft12(\Phi_{(S)})_{M_1\cdots M_{2n}} \nabla_L\nabla^L (\Phi_{(S)})^{M_1\cdots M_{2n}} \nn \\
&&\qquad-\ft12\left(M_\Phi^2+\frac{(K+S)(K-S)}{2L^2}\right)
(\Phi_{(S)})_{M_1\cdots M_{2n}}(\Phi_{(S)})^{M_1\cdots M_{2n}}+\cdots\biggr],
\eea
where the notation $\Phi_{(S)}$ indicates that there are $S$ indices along the
AdS$_5$ directions.

The above Lagrangian results in the AdS$_5$ equation of motion
\begin{equation}
[\nabla_\mu\nabla^\mu-\mu^2]\Phi_{(k)}=0,
\label{eq:mass1}
\end{equation}
where the effective mass is
\begin{equation}
\mu^2=M_\Phi^2+\fft{(K+S)(K-S)}{2L^2}-\nabla_m\nabla^m.
\label{eq:mass2}
\end{equation}
The angular Laplacian on $S^5$ gives rise to an angular momentum contribution.
While this is straightforward for scalar harmonics (and gives $J(J+4)$ in
the angular momentum $(J,0,0)$ representation), the spectrum is somewhat
more involved for general tensor harmonics.  We focus on a subset
of decompositions possible, namely when the field can be written in terms
of a symmetric transverse traceless ($ST^2$ henceforth) part on AdS times a
$ST^2$ part on
the sphere.  Sphere tensor harmonics of this form were discussed in
\cite{Rubin:1983be} but also see \cite{Higuchi:1986wu,Bianchi:2003wx}.
In particular, we need the eigenvalues of the d'Alembertian operator on
the sphere and in \cite{Rubin:1983be} these were found for general
$ST^2$ spherical harmonics.  It was also shown that the quadratic
Casimir operator of the $O(n_s+1)$ ($n_s=$ dimension of sphere)
for these harmonics was that of the $(J,K,0,\cdots 0)$ representation (these
label the length of each row in the Young tableau, and so $J\geq K$).
It was shown that for the first and second rank $ST^2$ harmonics ($K=1,2$)
that this is indeed the representation.  We take this to be the case for
the higher rank harmonics as well.
In any case, the eigenvalue of the d'Alembertian acting on these harmonics was
found to be
\be
\nabla^m \nabla_m = -\frac{J\left(J+n_s-1\right)-K}{L^2},
\ee
and $J>K$ as above.
This gives
\be
\mu^2=M_\Phi^2+\fft{(K+S)(K-S)}{2L^2}+\frac{J\left(J+4\right)-K}{L^2}.
\label{eq:mass2harm}
\ee
We stress here that while it is tempting to think
of $J=K$ as the intrinsic angular momentum and the `extra' coming from
$J\neq K$ as the `orbital' angular momentum, this is not a meaningful
concept on the sphere (as $E_0$ is the only meaningful `energy' on AdS).  The
charges are simply $(J,K,0)$ which label the representation.  The analog
of this in AdS will be addressed next.

\subsection{Energy in AdS$_5$}

To relate this effective mass to energy in AdS$_5$, we recall that states
in AdS$_5\times S^5$ may be labeled by their $SU(2,2|4)\supset SO(2,4)\times
SO(6)\supset SO(2)\times [SU(2)\times SU(2)]\times SO(6)$ quantum numbers
$(E,S_1,S_2; J_1,J_2,J_3)$ with $S_1$ and $S_2$ corresponding to spacetime
angular momentum and $J_1$, $J_2$, $J_3$ corresponding to the three commuting
angular momenta ($R$-charges) on the sphere.  Unitary representations in
AdS$_5$ are then labeled by $D(E_0,S_1,S_2;J_1,J_2,J_3)$ with $E_0$ the lowest
energy eigenvalue and $(J_1,J_2,J_3)$ are Dynkin labels for the $SO(6)$
representation.  Using $SU(2)\times SU(2)$ notation, the symmetric tracefree
field $\Phi_{(k)}$ corresponds to $D(E_0,S/2,S/2;J,K,0)$ where $S$ may be
thought of as conventional `angular momentum', and $J\geq K=2n-S$ label the
symmetric representation for `$R$-charge $(J,K,0)$' on the sphere.

Mass in Anti-de Sitter (or any curved) space is a somewhat ambiguous concept,
as it does not directly correspond to a Casimir of the symmetry algebra.
Nevertheless, we choose the standard convention where masslessness corresponds
to the propagation of a reduced number of helicity states.  In this case,
for symmetric rank-$S$ tensors in AdS$_{d+1}$ we have
\cite{Met2,Met3,Deser:2001pe,Deser:2001us,Deser:2001xr,Met,Deser:2003gw}
\begin{equation}
[\nabla_\mu\nabla^\mu-m^2+(2(d-2)-(d-5)S-S^2)]\Phi_{(S)}=0,
\label{eq:mass3}
\end{equation}
along with transverse-tracelessness, $\nabla^\mu\Phi_{\mu\cdots}=0$ and
$\Phi^\mu{}_{\mu\cdots}=0$.  This equation is then compared to the
eigenvalues of the d'Alembertian operator on AdS via the equation
\be
[\nabla_\mu\nabla^\mu-E_0(E_0-d)+S]\Phi_{(S)}=0.
\ee
The value of $E_0$ corresponding to this
definition of mass is then simply \cite{Met}
\begin{equation}
E_0=\fft{d}2+\sqrt{m^2L^2+\left(\fft{d}2+S-2\right)^2}.
\end{equation}
Comparing (\ref{eq:mass1}) and (\ref{eq:mass3}), we determine
\begin{equation}
E_0=\fft{d}2+\sqrt{\mu^2L^2+(d/2)^2+S},
\end{equation}
resulting in the AdS$_5$ ($d=4$) energy relation
\be
E_0=2+\sqrt{2\sqrt\lambda(K+S-2)+\ft12(K+S-2)(K-S)+J(J+4)+4+
\mathcal O(1/\sqrt\lambda)},
\ee
where we have introduced the usual definition $\sqrt{\lambda}=L^2/\alpha'$.
In this context, we wish to go as close to `$s$-wave' as possible, and
so we take $J=K$, its minimum value.
In the limit of large spin  or $R$-charge, but still in the short string
limit, $1\ll K+S\ll\sqrt\lambda$, we may expand $E_0$ to give
\be
E_0 \approx \sqrt{2\sqrt{\lambda}(K+S)}
\left(1+\frac{K-S}{8\sqrt{\lambda}}+\frac{K(K+4)}{2(K+S)\sqrt{\lambda}}+\cdots\right).
\label{eq:expres}
\ee

A comparable expression was derived in \cite{Frolov:2002av} for a
semi-classical string spinning in AdS$_5$ and carrying {\it only} orbital angular
momentum on $S^5$.  For vanishing orbital angular momentum (which corresponds
to the above discussion), the short-string result was given as
\cite{Frolov:2002av}
\begin{equation}
E\approx\sqrt{2\sqrt\lambda S}\left(1+\fft{S}{\sqrt\lambda}+\cdots\right),
\end{equation}
which differs from the $K=0$ limit of our result, (\ref{eq:expres}).  It is
not clear to us where this discrepancy arises from.  However, it is interesting
to note that the deviation from flat-space Regge behavior differs in sign
(and not just by a factor) between these two results.  One important difference to note
is that
for us the wave function of the string is spread over the entire sphere.  In \cite{Frolov:2002av}
the string is placed at a single {\it point} on the sphere, and so it may be that
this state corresponds to a large sum of KK states on the sphere.  For example,
the wave function being a delta function at the north pole (for scalar harmonics)
is given by
$\Sigma C^{(2)}_{\ell}(\cos(\theta))$, where $C^{(2)}_{\ell}(x)$ are the Gegenbauer (ultraspherical)
polynomials on $S^5$.  One should do likewise for placing the string
at the center of AdS. Using the resulting average $E_0=E_0(\langle
{\mathcal O}\rangle)$ 
may allow for more direct comparison
with results from \cite{Frolov:2002av}.

{}From our above computation of mass on AdS$_5$, we may also consider
strings with angular momentum, but not spin, on the sphere ({\it i.e.}~$K=0$).
In this case, $-\nabla_m\nabla^m$ has eigenvalue $J(J+4)$ for scalar
harmonics on $S^5$.  The resulting expression for $E_0$ is
\begin{eqnarray}
E_0&=&2+\sqrt{2\sqrt\lambda(S-2)+J(J+4)
-\ft12S(S-2)+4+\mathcal O(1/\sqrt\lambda)}\nonumber\\
&\approx&\sqrt{2\sqrt\lambda S+J^2-\ft12S^2},
\end{eqnarray}
in agreement with the leading behavior given in \cite{Frolov:2002av}.
Note that, in general, that the Kaluza-Klein spectrum of massive string
fields was investigated in \cite{Bianchi:2003wx}, based on the identification
of higher spin symmetry in the $\lambda\to0$ limit.  For leading Regge
states, the resulting spectrum was simply $E_0=2\ell+n$ where $\ell$ is the
string level and $n$ denotes the Kaluza-Klein harmonic on $S^5$.
As this is outside the scope of the short string case that we investigate
here, the results do not appear to be directly comparable.

\subsection{Conclusions}

Although the validity of this perturbative approach depends on taking
$K+S\ll\sqrt\lambda$, correspond to short strings, we feel the results
are robust within this limit.  Although the string effective action can
only be determined up to terms vanishing on-shell, this is all that is
required for the Freund-Rubin background.  Note, however, that we do not
prove that AdS$_5\times S^5$ is a consistent background for string
propagation, although this has been shown at the perturbative level
in the pure spinor formalism \cite{Berkovits:2004xu}.

We also wish to point out that the gravitational quadrupole interaction of
massive string states was previously investigated in
\cite{Giannakis:1998wi}, where the effective Lagrangian
(\ref{3ptlag}) was obtained as an intermediate step towards computing the
gravitational `$h$-factor' $h=S_LS_R/2S(S-1)$ for NSNS higher spin states.
Here the ten-dimensional spin contributions $S_L$ and $S_R$ arise from the
left and right movers, respectively, and the $h$ factor is defined by
the coupling term
\begin{equation}
\left(\nabla_\mu\nabla^\mu-m^2+hR_{\mu\nu\lambda\sigma}\ft12\Sigma^{\mu\nu}
\ft12\Sigma^{\lambda\sigma}\right)\Phi+\cdots=0,
\end{equation}
where $\Sigma^{\mu\nu}$ are Lorentz generators in the spin-$S$ representation.
As shown in \cite{Cucchieri:1994tx}, tree-level unitarity is violated for
gravitating states with $h\ne1$, and in particular the work of
\cite{Giannakis:1998wi} demonstrates this to be the case for massive string
states.  While this is not a serious issue for the full string theory, as it
merely indicates that the scattering becomes strongly coupled at the string
scale, it may nevertheless signify a breakdown in the effective field theory
description of the massive state $\Phi$, despite the indications of a
narrow width \cite{Iengo:2002tf,Iengo:2003ct}.

Curiously, we note that, while much effort was expended on computing the
$\Phi$-$F_{(5)}$-$F_{(5)}$-$\Phi$ contact terms the previous section, in
the end they do not contribute at all to the mass shift in a maximally
symmetric background.  This is because of the particular kinematical
combination of $F_{(5)}^2$ which enters (\ref{FinLag})
\begin{equation}
F^{MN\cdots}F^{PQ}{}_{\cdots}+\ft14F^{M\cdots}F^P{}_{\cdots}
\equiv\mathcal T^{MNPQ}
+\ft14 g^{PQ}\mathcal T^{MLN}{}_L.
\end{equation}
This vanishes identically for a symmetrical tensor $\mathcal T_{MNPQ}
\sim g_{MP}g_{NQ}-g_{MQ}g_{NP}$ contracted between symmetric tracefree
fields $\Phi$.  From this point of view, the effects of a closed
string spinning in AdS$_5\times S^5$ are due entirely to its interactions
with the NSNS background, and (at least to this order) the Ramond-Ramond
background can be completely ignored.  While we are uncertain as to whether
this holds at all perturbative orders, it nevertheless suggests the
possibility that semi-classical bosonic sigma-model approaches to spinning
strings in fact do not pick up any corrections from the Ramond-Ramond
background.

Essentially, what we have explored from the world-sheet point of view is
the coupling of massive higher spin states to the Type IIB supergravity
multiplet (or at least the metric and five-form components).  In principle,
if a manifestly supersymmetric formalism such as a superfield description
were to exist, then one would presumably be able to determine all relevant
couplings by the supersymmetric completion of the (Riemann) curvature induced
terms.  However we are not aware of any such approaches pertaining to the
IIB massive higher spin fields in long representations of the superalgebra.

Nevertheless, the structure of the IIB supergravity multiplet clearly
demonstrates the relation between Riemann and $F_{(5)}^2$ terms.
For example, integrability of the IIB supercovariant derivative
\begin{equation}
\mathcal D_M=\left[\nabla_M+\fft{i}{16\cdot5!}F_{NPQRS}
\Gamma^{NPQRS}\Gamma_M+\cdots\right]\epsilon
\end{equation}
may be expressed as $[\mathcal D_M,\mathcal D_N]\epsilon
=\ft14\mathcal R_{MN}\epsilon$ where \cite{Papadopoulos:2003jk}
\begin{eqnarray}
\mathcal R_{MN}&=&\left[R_{MNPQ}-\fft1{48}F_{MPABC}F_{NQ}{}^{ABC}
\right]\Gamma^{PQ}\nonumber\\
&&+\fft1{24}\left[i\nabla_{[M}F_{N]PQRS}+\fft14F_{MNPAB}F_{QRS}{}^{AB}
-\fft14g_{P[M}F_{N]QABC}F_{QR}{}^{ABC}\right]\Gamma^{PQRS}\nonumber\\
&&+\cdots.
\end{eqnarray}
{}From this point of view, supersymmetry indicates that any term proportional
to Riemann will be related to similar terms containing both $F_{(5)}^2$
and $\nabla F_{(5)}$.  The latter are unimportant for our case, as covariant
derivatives of $F_{(5)}$ vanish for the maximally supersymmetric backgrounds
that we consider.  It would be interesting to see if this supercovariant
derivative, while originally acting on spinors, can be extended to provide
a means of constructing supersymmetric actions for higher-rank bosonic
fields $\Phi$.  This is perhaps most promising for the case of $p$-form
actions, as they may be constructed out of spinor bilinears.
These ideas may end up involving the geometry and hidden symmetries of
the supercovariant derivative, an idea which has attracted some recent
interest
\cite{Gauntlett:2002fz,Duff:2003ec,Hull:2003mf,Papadopoulos:2003pf,%
Gran:2005wu,Gran:2005ct}.

Finally, we note that, while it has not been extensively studied, it
is entirely feasible to work with Ramond-Ramond emission vertex operators
in the RNS formalism.  In this fashion, it is for example possible to
go beyond linearized superfields \cite{deHaro:2002vk} to explore the
supersymmetric completion of $R^4$ terms from the world-sheet point of view;
this was examined for $R^2(\nabla F_{(5)})^2$ in \cite{Peeters:2003pv}.
Of course, a more direct approach to Ramond-Ramond backgrounds may
nevertheless lie within a Green-Schwarz
\cite{Metsaev:1998it,Metsaev:2000yf}
or pure spinor formulation
\cite{Berkovits:2000fe,Berkovits:2000wm,Berkovits:2000yr,Berkovits:2002qx,%
Berkovits:2004xu},
or ultimately within string field theory.

\section*{Acknowledgments}
This project came about following early discussions with I.~Giannakis and
M.~Porrati about the feasibility of extending the string gravitational
quadrupole coupling result of \cite{Giannakis:1998wi} to include interactions
with a Ramond-Ramond background.  We wish to thank J. Davis, F.~Larsen and
D.~Vaman for useful comments and helpful advice.  B.B. is especially grateful
for conversations with P.~de Medeiros.
This work was supported in part by the
US~Department of Energy under grant DE-FG02-95ER40899.
%%%%%%%%%%%%%%%%%%%%%%%%%%%%%%

\end{document}